# The Photodetection Process under the Conditions of an Imperfect Measurement Device

[1]Miroshnichenko G.P., [2]Trifanov A.I.

Saint-Petersburg State University of Information Technologies, Mechanics and Optics

49 Kronverksky Avenue, Saint Petersburg, Russia 192101

[1]gpmirosh@gmail.com,
[2]alextrifanov@gmail.com

**Abstract:** The process of cavity mode quantum state photodetection subject to a nonideal measurement device is under consideration. A set of nonorthogonal probabilistic operator valued measures (POVM's) describing the photodetection process is suggested. The superoperators of conditional system evolution and their Kraus representations are obtained. Developed formalism is used to the quantum state reconstruction problem. We recalculate the probabilities of hypotheses about the initial state of cavity quantum mode in the case of unknown superposition of vacuum and single photon states.

## 1. Introduction

Quantum measurement [1,2] is one of the fundamental problems in modern physics. It plays a significant role in different applications of quantum science. Nowadays the quantum state tomography [3,4], quantum process tomography [5,6] and detector POVM's reconstruction [7] are the objects of interest. A measurement device is the essential part of the quantum computer and quantum communication systems [8,9].

Photonic technologies [10] seem to be the natural candidate for the realization of the quantum information transmission. Here the photodetection process is used to access the data about the state of the transmission channel. The theoretical basis of the electromagnetic (EM) field measurement description was formed in [11,12,13]. Further development one can find in [14,15,16]. The discreet [17, 18] and continuous [19, 20, 21, 22] process for "internal" [23] and "external" [24] (dependent on location of the photodetector) models have been discussed. Quantum jump superoperators of Srinivas-Davies (SD) [14] and E-models [25] have been constructed and thoroughly investigated [21,22] under the conditions of the imperfect measurement apparatus [26].

In the discreet model (see also works on one atom maser [27, 28]) methods of statistical properties of the EM field [29] and their appendixes to some protocols of the quantum measurement in information technologies [30] have been developed. In this measurement protocol the state of the cavity mode is tested with the help of a two-level atom-pointer, which passes through the cavity and "collects" information about the state of the quantum mode. After the interaction the atom undergoes selective measurement in an ionization chamber. Dependent on the state in which the atom is prepared there are two operation modes of the measurement device – a photon counter and a quantum counter. The photon counter is realized when the atom is prepared in its ground state, whereas the quantum counter is realized in the case of the initially excited atom. Information, fidelity and reversibility of these two types of measurement were investigated in [31].

Here we are investigating the photodetection process which is realized in the discreet model as a photon counter subject to nonideality originates from atom state detection in ionizing chamber. This imperfection implies the possibility to miss the atom as well as to get incorrect information about its state. Probabilities of such events may be obtained directly from the experiment. For instance, one can calculate the frequency of the detector's clicks which tune to measure the atomic ground state for the atom prepared in excited state and vice a versa. Such errors are originated from the interaction between the atom and the pulses of the ionizing classical electromagnetic (EM) field inside the chamber. These pulses may change the population of the atomic energy levels without photoionization. This population change leads to erroneous statistics of photocounts.

This work is organized as follows: in section 2 the POVMs for detecting the atomic state are introduced. The correspondent superoperators and their Kraus representations subject to imperfection are suggested. In section 3 we construct the POVMs on the cavity quantum field states space and propose evolution superoperators and their corresponding Kraus representations. Section 4 contains the simple example of developed formalism in the case of the single photon cavity mode detection. The calculation of a posteriori probabilities of the hypotheses about the initial state of the cavity field is executed with the help of Bayes' theorem. Section 5 contains the discussion and concludes the article.

**2. The parameterization of the photodetector**

Let's consider the atom state measurement using ionizing chamber. We are interested in the two-level pointer (atom) state, which is the observable in our experiment: ground state $|g\rangle$ or excited state $|e\rangle$. The complete set of spectral projectors $\mathrm{P}_j^A$, $j \in \{e, g\}$ associated with the correspondent operator of the observable satisfies the following condition:

$$\mathrm{P}_g^A + \mathrm{P}_e^A = I_A, \tag{1}$$

where $I_A$ is the identity operator, and projectors $\mathrm{P}_j$ are defined as follows:

$$\mathrm{P}_g^A = |g\rangle\langle g|, \; \mathrm{P}_e^A = |e\rangle\langle e|. \tag{2}$$

The result of the observable measurement may be characterized with the help of the random variable $\xi$ which can take on a three possible values: $\xi = 1$ correspondent to the case when atom is "seen" by the detector of the ionization chamber tuned to the ground state $|g\rangle$ detection; the case $\xi = 2$ corresponds to the click of the detector tuned to the excited state $|e\rangle$ detection; $\xi = 0$ corresponds to the atom passing through the chamber without any detector's click. The complete set of POVMs $\Pi_\xi$, $\xi = 0, 1, 2$, may be associated with this random variable:

$$\Pi_0^A + \Pi_1^A + \Pi_2^A = I_A. \tag{3}$$

Here

$$\Pi_1^A + \Pi_2^A = \varepsilon_g |g\rangle\langle g| + \varepsilon_e |e\rangle\langle e|, \tag{4}$$

$$\Pi_0^A = (1 - \varepsilon_g)|g\rangle\langle g| + (1 - \varepsilon_e)|e\rangle\langle e|. \tag{5}$$

Quantities $\varepsilon_g$ and $\varepsilon_e$ determine the quantum efficiencies of detectors. Lets $\rho_A$ is density matrix of the atom just before the measurement. Then the probability of the experiment outcome characterizing by the random variable $\xi$ is

$$P(\xi) = Tr_A\left[\rho_A \Pi_\xi^A\right]. \tag{6}$$

Here $Tr_A$ is the trace operation in the atomic state space. Quantum operation $\Lambda_\xi$ correspondent to the experiment outcome $\xi$ is introduced in the following way:

$$Tr_A\left[\Lambda_\xi \rho_A\right] = Tr_A\left[\rho_A \Pi_\xi^A\right], \tag{7}$$

$$\Lambda_\xi = \sum_k M_{\xi,k} M_{\xi,k}^\dagger, \tag{8}$$

where $M_{\xi,k}$ is the set of bounded operators (state transformers) and

$$\sum_k M_{\xi,k}^\dagger M_{\xi,k} = \Lambda_\xi^* I = \Pi_\xi^A \tag{9}$$

Here $\Lambda_\xi^*$ is a conjugated superoperator. Now let us take into account one more kind of the nonideality associated with the case when the result of the state detection of the atom prepared in its ground state is described by value $\xi = 2$, and vice versa. For this purpose we introduce corresponding conditional probabilities $P(\xi = 1|e) = |\alpha_{1,e}|^2$ and $P(\xi = 2|g) = |\alpha_{2,g}|^2$ which may be determined empirically. Let us write the expression for the total probability of the random variable $\xi$:

$$P(\xi) = P(\xi|g)P(g) + P(\xi|e)P(e) \tag{10}$$

where $P(g) = Tr_A\left[\rho_A P_g^A\right]$, $P(e) = Tr_A\left[\rho_A P_e^A\right]$. Using (4), (6) and (10) we obtain the expressions for the atomic POVMs:

$$\Pi_1^A = |\alpha_{1,g}|^2 |g\rangle\langle g| + |\alpha_{1,e}|^2 |e\rangle\langle e|, \tag{11}$$

$$\Pi_2^A = |\alpha_{2,e}|^2 |e\rangle\langle e| + |\alpha_{2,g}|^2 |g\rangle\langle g|. \tag{12}$$

Here $|\alpha_{1,g}|^2 + |\alpha_{2,g}|^2 = \varepsilon_g$ and $|\alpha_{1,e}|^2 + |\alpha_{2,e}|^2 = \varepsilon_e$. In the case $\xi = 0$ it is convenient to write $|\alpha_{0,g}|^2 = 1 - \varepsilon_g$, and $|\alpha_{0,e}|^2 = 1 - \varepsilon_e$. Using (9) operators $\Pi_\xi$ may be represented in the form

$$\Pi_\xi^A = \sum_{\mu,\nu \in \{g,e\}} M_{\xi,\mu\nu}^\dagger M_{\xi,\mu\nu} \tag{13}$$

Each of the operators $M_{\xi,\mu\nu}$ in Eq. (13) specifies the certain process inside the ionization chamber: $M_{\xi,gg} = \alpha_{\xi,gg}|g\rangle\langle g|$ corresponds to the case when atom is prepared in state $|g\rangle$ causing the click of $\xi$-th detector ($\xi = 1,2$); similarly, $M_{\xi,ee} = \alpha_{\xi,ee}|e\rangle\langle e|$ corresponds to the case when atom is prepared in state $|e\rangle$ causing the click of $\xi$-th detector; $M_{\xi,ge} = \alpha_{\xi,ge}|e\rangle\langle g|$ describes the transition $|g\rangle \to |e\rangle$ under the classical field pulse followed by $\xi$-th detector click; and $M_{\xi,eg} = \alpha_{\xi,eg}|g\rangle\langle e|$ - describes the transition $|e\rangle \to |g\rangle$ under the classical field pulse followed by $\xi$-th detector click. The meaning of the terms $M_{\xi,\mu\nu}$ in the case of $\xi = 0$ is: $M_{0,gg}$ - the atom prepared in state $|g\rangle$ has passed through the ionization chamber without any click; $M_{0,ee}$ - the atom prepared in state $|e\rangle$ has passed through the ionization chamber without any click; finally $M_{0,eg}$ and $M_{0,ge}$ - the atom is not detected, but it interacts with the internal fields of the chamber followed by the transitions $|g\rangle \to |e\rangle$ and $|e\rangle \to |g\rangle$ correspondingly. From Eqs. (11) and (12) one can obtain:

$$|\alpha_{\xi,gg}|^2 + |\alpha_{\xi,ge}|^2 = |\alpha_{\xi,g}|^2, \quad |\alpha_{\xi,ee}|^2 + |\alpha_{\xi,eg}|^2 = |\alpha_{\xi,e}|^2 \tag{14}$$

Expression for $P(\xi)$ in form (6) may be written as follows:

$$P(\xi) = Tr_A\left[\rho_A \Pi_\xi^A\right] = Tr_A\left[\sum_{\mu,\nu \in \{g,e\}} M_{\xi,\mu\nu} \rho_A M_{\xi,\mu\nu}^\dagger\right]. \tag{15}$$

It should be noticed that in Eq. (13) one may take into account the processes like $|e\rangle\langle g|\langle e|\langle g|$ but the probabilities of such series of repopulations are negligible compared to the process described above.

### 3. The process of photodetection

In this section we'll consider the process of the cavity quantum electromagnetic field detection with a two-level atom-pointer in use. Let the state of the atom–field system at the moment when the atom is entering the ionization chamber be described by the joint density matrix $\rho_{AF}$. The measurement (due to the Neuman postulate) reduces the atomic state and the final state of the atom-field system is described by the following density operator:

$$\tilde{\rho}_{AF} = \sum_{\xi=0}^{2}(\Lambda_\xi \otimes I_F)\rho_{AF} = \sum_{\xi=0}^{2}\sum_{\mu,\nu \in \{g,e\}} M_{\xi,\mu\nu} \rho_{AF} M_{\xi,\mu\nu}^\dagger \tag{16}$$

Ignoring the result of the detection one can obtain the reduced density matrix of the entire ensemble, taking the trace in the atomic state space

$$\rho_F = Tr_A[\tilde{\rho}_{AF}]. \tag{17}$$

Additional information about the state of the atom leaving the resonator may be used for the entire ensemble decomposition. Consequently one can obtain the subensembles, each of them corresponds to one of the detection results $\xi$:

$$\tilde{\rho}_F^{[\xi]} = Tr_A\left[\sum_{\mu,\nu \in \{g,e\}} M_{\xi,\mu\nu} \rho_{AF} M_{\xi,\mu\nu}^\dagger\right]. \tag{18}$$

The probability of this subensemble may be found as follows:

$$P(\xi) = Tr_F\left[\tilde{\rho}_F^{[\xi]}\right] = Tr_F\left[\tilde{\rho}_F^{[\xi]} I_F\right]. \tag{19}$$

Using the decomposition like in Eq.(3) one can perform the identity resolution in the space of the field states thereby POVMs $\Pi_\xi^F$ and superoperators $\Xi_\xi$ will be determined. For the certainty we accept that the atom just before the interaction with the intracavity EM field had been prepared in its ground state $\rho_A = |g\rangle\langle g|$. We assign the arbitrary density operator $\rho_F$ to the initial state of the resonator mode. Let us neglect the relaxation processes in the system and examine the situation of the unitary evolution. Then system state $\rho_{AF}$ right after the interaction and just before the detection may be written in the following way:

$$\rho_{AF} = U_\tau \rho_A \otimes \rho_F U_\tau^\dagger, \qquad (20)$$

where $U_\tau$ is the unitary evolution operator of the atom-field system and $\tau$ is the interaction time. We implement the identity decomposition in the field state space. Let as introduce the superoperators $\Xi_\xi$:

$$Tr_F\left[\Xi_\xi \rho_F\right] = Tr_F\left[\tilde{\rho}_F^{[\xi]} I_F\right] = Tr_{AF}\left[\sum_{\mu,\nu \in \{g,e\}} M_{\xi,\mu\nu} \rho_{AF} M_{\xi,\mu\nu}^\dagger\right]. \qquad (21)$$

Then using Eq. (20) we define the Kraus operators (transformers) $K$:

$$K_{\xi,\mu\nu} = \sum_{\eta \in \{g,e\}} \langle \eta | M_{\xi,\mu\nu} U_\tau | g \rangle. \qquad (22)$$

Subject to equations for $M_{\xi,\mu\nu}$ the operators $K_{\xi,\mu\nu}$ may be expressed in the following way:

$$K_{\xi,\mu\nu} = \sum_{\eta \in \{g,e\}} \alpha_{\xi,\mu\nu} \langle \eta | \nu \rangle \langle \mu | U_\tau | g \rangle = \sum_{\eta \in \{g,e\}} \alpha_{\xi,\mu\nu} \delta_{\nu\eta} U_{\mu g}. \qquad (23)$$

Here $\delta_{\nu\eta}$ is Kronecker delta, $U_{gg} = \langle g | U_\tau | g \rangle$ and $U_{eg} = \langle e | U_\tau | g \rangle$. From (23) it is simple to obtain the expression for the superoperators $\Xi_\xi$:

$$\Xi_\xi = \sum_{\mu,\nu \in \{g,e\}} K_{\xi,\mu\nu} K_{\xi,\mu\nu}^\dagger = |\alpha_{\xi,g}|^2 U_{gg} U_{gg}^\dagger + |\alpha_{\xi,e}|^2 U_{eg} U_{eg}^\dagger. \qquad (24)$$

We conclude from here that the "quantum jumps", which may take place during the atom state measurement, don't contribute to the detection result (this was pointed out in [27] in connection with the one-atom maser investigation). Consequently they may be dropped in the calculation of allowance due to the nonideality. Process like $M_{\xi,ge} = \alpha_{\xi,ge} |e\rangle\langle g|$ may be useful for the calculation of $\alpha$ (which is introduced phenomenologically here) from the origin principles.

**4. The model problem: Bayes' theorem**

Here we illustrate the application of the formalism developed above. For instance, let us calculate a posteriori probability of the hypothesis about the state of the intracavity field. Suppose that the initial pure state of the field is the linear combination of the vacuum and single photon states. $\rho_F = |\psi\rangle\langle\psi|$ is its density matrix which may be parameterized as follows:

$$|\psi\rangle = |\psi(\theta)\rangle = \cos\frac{\theta}{2}|0\rangle + \sin\frac{\theta}{2}|1\rangle, \quad \theta \in [0,\pi] \qquad (25)$$

Using Eq.'s (19), (21) one can write the expression of the total probability of the random variable $\xi = 1,2$. This probability depends on the state of the cavity field $\rho_F = \rho_F(\theta)$:

$$P(\xi|\theta) = Tr_F\left[\Xi_\xi \rho_F\right] = P(\xi|g) P(g|\theta) + P(\xi|e) P(e|\theta), \qquad (26)$$

where $P(\xi|g)$ and $P(\xi|e)$ are defined above, and

$$P(g|\theta) = Tr_F\left[\rho_F \mathrm{P}_g^F\right], P(e|\theta) = Tr_F\left[\rho_F \mathrm{P}_e^F\right]. \tag{27}$$

Here $\mathrm{P}_g^F$ and $\mathrm{P}_e^F$ are spectral projectors in the field state space, which correspond to projectors $\mathrm{P}_g^A$ and $\mathrm{P}_e^A$ in the atomic state space. After the measurement one of the possible values of $\xi$ is realized and the probability of the corresponding hypotheses $P(\theta)$ about the initial state may be recalculated in the following way (Bayes' theorem):

$$P(\theta|\xi) = \frac{P(\xi|\theta)P(\theta)}{\sum_\theta P(\xi|\theta)P(\theta)}. \tag{28}$$

In our case this expression becomes

$$P(\theta|1) = \frac{\left\{|\alpha_{\xi,g}|^2 Tr_F\left[U_{gg}\rho_F U_{gg}^\dagger\right] + |\alpha_{\xi,e}|^2 Tr_F\left[U_{eg}\rho_F U_{eg}^\dagger\right]\right\}P(\theta)}{\sum_\theta \left\{|\alpha_{\xi,g}|^2 Tr_F\left[U_{gg}\rho_F U_{gg}^\dagger\right] + |\alpha_{\xi,e}|^2 Tr_F\left[U_{eg}\rho_F U_{eg}^\dagger\right]\right\}P(\theta)}. \tag{29}$$

Using explicit formulas for operators $U_{gg}$ и $U_{eg}$:

$$U_{gg} = \cos\left(\Omega\tau\sqrt{a^\dagger a}\right), U_{eg} = -\frac{i}{\sqrt{aa^\dagger}}\sin\left(\Omega\tau\sqrt{aa^\dagger}\right)a, \tag{30}$$

where $\Omega$ - single photon Rabi frequency, one can obtain:

$$P(\theta|\xi) = \frac{1}{\pi}\left[1 + \frac{\left(|\alpha_{\xi,g}|^2 - |\alpha_{\xi,e}|^2\right)\sin^2(\Omega\tau)\cos(\theta)}{2|\alpha_{\xi,g}|^2 + \left(|\alpha_{\xi,e}|^2 - |\alpha_{\xi,g}|^2\right)\sin^2(\Omega\tau)}\right]. \tag{31}$$

Particularly, for the case $\xi = 0$ we get:

$$P(\theta|0) = \frac{1}{\pi}\left[1 + \frac{(\varepsilon_e - \varepsilon_g)\sin^2(\Omega\tau)\cos(\theta)}{2(1-\varepsilon_g) + (\varepsilon_g - \varepsilon_e)\sin^2(\Omega\tau)}\right]. \tag{32}$$

On Fig. 1 the probability density $P(\theta|\xi)$ of the recalculated hypotheses is depicted for the cases $\xi = 0$ and $\xi = 1$. Here we examine the case when $P(\theta)$ initial uniformly distributed on the interval $[0, \pi]$. For $\theta = 0$ (Fig. 2) we graph the probability density $P(0|0)$ as a function of the first detector efficiency $\varepsilon_g$ under $\varepsilon_e$ fixed.

## 5. Discussion and conclusion

We considered the model of the imperfect measurement process of the quantum cavity mode. This model includes the possibility of missing the atom-pointer and detection errors associated with the interaction process which takes place in the ionizing chamber. The set of detectors POVMs allowing all types of nonideality under the consideration is suggested. Using this set we construct the atomic superoperators corresponding to each of the measurement outcome. The Kraus representation for each superoperator is suggested. Further this model of the atom-pointer measurement device is used for testing the state of the cavity quantum field. Here for simplicity we suppose that the time interaction is shorter than the lifetime of the excited atomic energy level and the cavity decay rate. This allows us to consider the case of unitary evolution of combining atom-field system. Under these simplifications we obtain the Kraus representations for the superoperators which act in the field state space and then use them for the posterior recalculation probabilities of the hypotheses about the distribution of the initial field state (Bayes' theorem).

Besides the application described above, it would be interesting to study the effect of the nonideality on the estimation of statistical characteristics of the quantum mode as well as reversibility of the photodetection process. The photocount statistics for different photodetection regimes subject to the possibility of the atom missing in the presence of the relaxation process was studied in [29]. In [31] it was shown that in the ideal case of the photon counter the measurement process is irreversible. The existence of erroneous counts of the photodetector may influence the value of the detection reversibility.

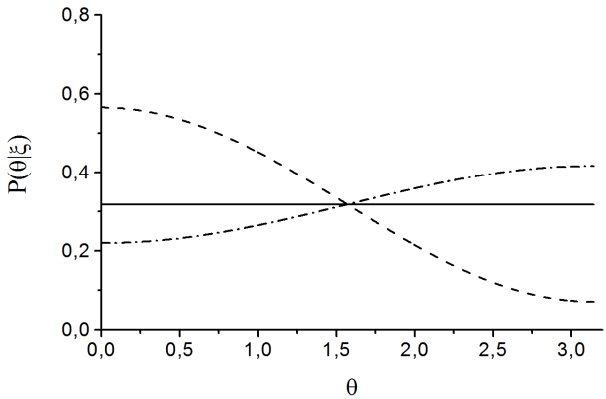
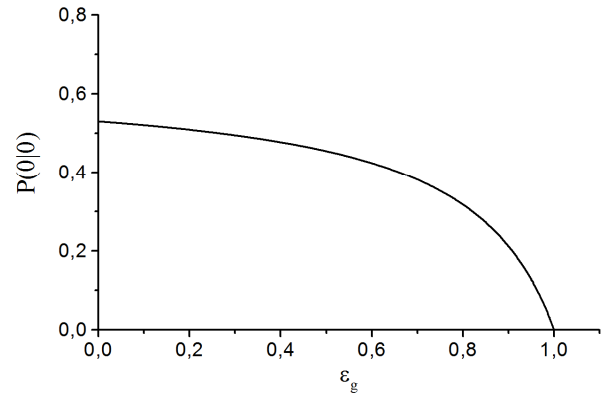

**Fig. 1** There calculated probability densities for the hypotheses: ($\Omega\tau = \pi/2$): $P(\theta|1)$, $|\alpha_{1,g}|^2 = 0.8$ (dotted line); $P(\theta|0)$, $\varepsilon_g = 0.9$, $\varepsilon_e = 0.8$ (dashed-dotted line); $P(\theta) = 1/\pi$ is a priori probability density (solid line).

**Fig. 2** Dependence $P(0|0)$ of the value $\varepsilon_g$ - quantum efficiency of the first detector under the value $\varepsilon_e = 0.8$ fixed; $\Omega\tau = \pi/2$, $\theta = 0$.

**Acknowledgements**

The work is partially supported by FTP "Scientific and Pedagogical Personnel of Innovative Russia" for 2009–2013 (contracts P689 NK-526P, 14.740.11.0879 and 16.740.11.0030), FTP "Investigation and engineering in priority lines of development scientific and technological complexes in Russia 2007-2013" (contract 07.514.11.4146) and Russian Foundation for Basic Research (grant 11-08-00267).